\documentclass{aa}

\usepackage{epsf}
\usepackage{graphics}

\begin{document}

\def\clipfig#1{\def\lbracket{[}\def\testit{#1}%
    \ifx\testit\lbracket\let\next=\optclipfig\else\let\next=\stdclipfig\fi%
    \next{#1}}
%
\newcommand {\hclipfig} [7] {\clipfig[#7]{#1}{#2}{#3}{#4}{#5}{#6}}
%
\def\usemodepsfig {\global\def\cfmode{x}\typeout{*** set clipfig to PSFIG mode ***}}
\def\usemodeepsf  {\global\def\cfmode{}\typeout{*** set clipfig to EPSF mode ***}}
\def\useunitmm    {\global\def\cfunit{x}\typeout{*** set clipfig to use mm as unit ***}}
\def\useunitcm    {\global\def\cfunit{}\typeout{*** set clipfig to use cm as unit ***}}
\def\clipfigsettings {\ifx\cfmode\empty\def\ccfmode{EPSF }\else\def\ccfmode{PSFIG }\fi%
    \ifx\cfunit\empty\def\ccfunit{cm }\else\def\ccfunit{mm }\fi%
    \typeout{*** current clipfig settings: \ccfmode mode, using \ccfunit as unit ***}}
%
%
%
%
\def\stdclipfig#1#2#3#4#5#6{\ifx\cfmode\empty%
    \let\next=\eclipfig\else\let\next=\pclipfig\fi%
    \next{#1}{#2}{#3}{#4}{#5}{#6}}
\def\optclipfig#1#2]#3#4#5#6#7#8{\ifx\cfmode\empty%
    \let\next=\ehclipfig\else\let\next=\phclipfig\fi%
    \next{#3}{#4}{#5}{#6}{#7}{#8}{#2}}
%
%
%
\newcommand {\pclipfig}[6] {\ifx\cfunit\empty%
        \psfig{figure=#1.ps,width=#2cm,bbllx=#3cm,bblly=#4cm,bburx=#5cm,%
           bbury=#6cm,clip=}\else%
        \psfig{figure=#1.ps,width=#2mm,bbllx=#3mm,bblly=#4mm,bburx=#5mm,%
           bbury=#6mm,clip=}\fi}
\newcommand {\phclipfig}[7] {\ifx\cfunit\empty%
        \hspace{#7cm}\psfig{figure=#1.ps,width=#2cm,bbllx=#3cm,bblly=#4cm,%
           bburx=#5cm,bbury=#6cm,clip=}\else%
        \hspace{#7mm}\psfig{figure=#1.ps,width=#2mm,bbllx=#3mm,bblly=#4mm,%
           bburx=#5mm,bbury=#6mm,clip=}\fi}
%
%
%
\newcommand {\eclipfig}[6]{%
  \ifx\cfunit\empty\epsfxsize=#2cm\else\epsfxsize=#2mm\fi%
  \epsfclipon\epsfverbosetrue%
  \cfcmtopspts{#3}\cfllxi=\cftempi\cfllxf=\cftempf%
  \cfcmtopspts{#4}\cfllyi=\cftempi\cfllyf=\cftempf%
  \cfcmtopspts{#5}\cfurxi=\cftempi\cfurxf=\cftempf%
  \cfcmtopspts{#6}\cfuryi=\cftempi\cfuryf=\cftempf%
  \def\cfstra{\number\cfllxi.\number\cfllxf}%
  \def\cfstrb{\number\cfllyi.\number\cfllyf}%
  \def\cfstrc{\number\cfurxi.\number\cfurxf}%
  \def\cfstrd{\number\cfuryi.\number\cfuryf}%
  \hbox{\epsfbox[{\cfstra} {\cfstrb} {\cfstrc} {\cfstrd}]{#1.ps}}}
\newcommand {\ehclipfig}[7]{%
  \ifx\cfunit\empty\epsfxsize=#2cm\else\epsfxsize=#2mm\fi%
  \epsfclipon\epsfverbosetrue%
  \cfcmtopspts{#3}\cfllxi=\cftempi\cfllxf=\cftempf%
  \cfcmtopspts{#4}\cfllyi=\cftempi\cfllyf=\cftempf%
  \cfcmtopspts{#5}\cfurxi=\cftempi\cfurxf=\cftempf%
  \cfcmtopspts{#6}\cfuryi=\cftempi\cfuryf=\cftempf%
  \def\cfstra{\number\cfllxi.\number\cfllxf}%
  \def\cfstrb{\number\cfllyi.\number\cfllyf}%
  \def\cfstrc{\number\cfurxi.\number\cfurxf}%
  \def\cfstrd{\number\cfuryi.\number\cfuryf}%
  \ifx\cfunit\empty\hspace{#7cm}\else\hspace{#7mm}\fi%
  \hbox{\epsfbox[{\cfstra} {\cfstrb} {\cfstrc} {\cfstrd}]{#1.ps}}%
  \vspace{-1mm}}
%
%
%
\newdimen\cfllxi \newdimen\cfllyi  \newdimen\cfurxi  \newdimen\cfuryi
\newdimen\cfllxf \newdimen\cfllyf  \newdimen\cfurxf  \newdimen\cfuryf
\newdimen\cftemp \newdimen\cftempi \newdimen\cftempf
\newdimen\cfpspoint \cfpspoint=1bp
%
%
%
\newcommand{\cfcmtopspts}[1]{\ifx\cfunit\empty%
  \cftemp=#1cm\else\cftemp=#1mm\fi%
  \multiply\cftemp10\divide\cftemp\cfpspoint%
  \cftempf=\cftemp\divide\cftemp10\cftempi=\cftemp\multiply\cftemp10%
  \advance\cftempf-\cftemp}
%
%
\def\cfmode{}\def\cfunit{}\clipfigsettings
%

\useunitmm


\newcommand{\lb}{$\lambda$}
\newcommand{\sm}[1]{\footnotesize {#1}}
\newcommand{\inft}{$\infty$}
\newcommand{\vlv}{$\nu L_{\rm V}$}
\newcommand{\lv}{$L_{\rm V}$}
\newcommand{\lx}{$L_{\rm x}$}
\newcommand{\lsoft}{$L_{\rm 250eV}$}
\newcommand{\lhard}{$L_{\rm 1keV}$}
\newcommand{\vlsoft}{$\nu L_{\rm 250eV}$}
\newcommand{\vlhard}{$\nu L_{\rm 1keV}$}
\newcommand{\vlir}{$\nu L_{60\mu}$}
\newcommand{\ax}{$\alpha_{\rm x}$}
\newcommand{\aopt}{$\alpha_{\rm opt}$}
\newcommand{\aox}{$\alpha_{\rm ox}$}
\newcommand{\aoxh}{$\alpha_{\rm oxh}$}
\newcommand{\airhard}{$\alpha_{\rm 60\mu-hard}$}
\newcommand{\aoxsoft}{$\alpha_{\rm ox-soft}$}
\newcommand{\aio}{$\alpha_{\rm io}$}
\newcommand{\aixs}{$\alpha_{\rm ixs}$}
\newcommand{\aixh}{$\alpha_{\rm ixh}$}
\newcommand{\hb}{H$\beta$}
\newcommand{\nh}{$N_{\rm H}$}
\newcommand{\nhgal}{$N_{\rm H,gal}$}
\newcommand{\nhfit}{$N_{\rm H,fit}$}
\newcommand{\ale}{$\alpha_{\rm E}$}
\newcommand{\cts}{$\rm {cts\,s}^{-1}$}
\newcommand{\pl}{$\pm$}
\newcommand{\kev}{\rm keV}
\newcommand{\rb}[1]{\raisebox{1.5ex}[-1.5ex]{#1}}
\newcommand{\ten}[2]{#1\cdot 10^{#2}}
\newcommand{\msun}{$M_{\odot}$}
\newcommand{\dM}{\dot M}
\newcommand{\dMM}{$\dot{M}/M$}
\newcommand{\dMedd}{\dot M_{\rm Edd}}
\newcommand{\kms}{km\,$\rm s^{-1}$}


\title{RX J2217.9--5941: A highly X-ray variable Narrow-Line Seyfert1 galaxy
\thanks{bases in part on observations at the European Southern Observatory La
Silla (Chile) with the ESO/MPG 2.2m telescope, the ESO 1.52m telescope in 
August 1992 and September 1995, and the ESO 3.6m in January 2001}
}
\author{D. Grupe\inst{1},
H.-C. Thomas\inst{2},
\and K.M. Leighly\inst{3, 4}
}
\offprints{\\ D. Grupe (dgrupe@xray.mpe.mpg.de)}
\institute{MPI f\"ur extraterrestrische Physik, Postfach 1312, 
85741 Garching, Germany
\and MPI f\"ur Astrophysik, Karl-Schwarzschild-Str. 1, 85748 Garching, Germany
\and Columbia Astrophysics Laboratory, 538 West 120th St., New York, 
NY 10027, U.S.A.
\and Dept. of Physics and Astronomy, University of Oklahoma, 440 W. Brooks St.,
Norman, OK 73019, USA 
}
\date{received  26 October 2000; accepted 23 January 2001}

\abstract{
We report the discovery of a highly X-ray variable AGN, RX
J2217.9--5941.  This object was bright during the ROSAT All-Sky Survey
(RASS), during which a decrease in the count rate by a factor of 12
was observed.  It was found to be much fainter in follow-up HRI
observations and is therefore an X-ray transient AGN candidate.  On
long time scales of years, its count rate decreased by a factor of
about 30 between the RASS and the ROSAT HRI and ASCA observations in
1998.   Analysis of the ASCA data, complicated by source confusion and
poor statistics, 
tentatively indicates a steep spectrum in the faint state.  There is no
evidence for variability among 5 optical observations of the object
obtained between 1992 and 1998.  
We discuss the variability of RX J2217.9--5941 and its possible X-ray
transient nature.
\keywords{accretion, accretion disks -- galaxies: active -- galaxies: nuclei
-- galaxies: Seyfert
-- galaxies: individual (RX J2217.9--5941)}
}

\maketitle
\markboth{D. Grupe et al.: RX J2217.9--5941: A highly X-ray variable NLS1}{ }

\section{Introduction}

With the launch of the X-ray satellite ROSAT (Tr\"umper 1983) it had become
possible to study the X-ray spectra of  Active Galactic Nuclei
(AGN) in the energy band between 0.1 and
2.4 keV with the Position Sensitive Proportional Counter (PSPC, Pfeffermann et
al., 1986).
In this energy band AGN show typical variabilities by factors of 2-3 
on timescales of years
(e.g. Grupe et al. 2001).  However, some  AGN show rapid X-ray
variabilities by factors of more than a factor of ten on time scales of days. The most
extreme example of these persistent
rapid X-ray variable AGN is the Narrow-line Seyfert 1 galaxy (NLS1) 
IRAS 13224--3809 which has
shown flux variabilities of almost 60 in only two days and a shortest doubling timescale
of 800 s (Boller et al., 1997a).
Another example is the NLS1 PHL 1092 which has shown a
maximum intensity variability by a factor of $\approx$21 between Einstein and
ROSAT observations 
(Forster \& Halpern 1996) and by a factor of $\approx$14 during ROSAT
 High Resolution Imager (HRI) monitoring (Brandt et al. 1999).

The most extreme form of variability is 
 the X-ray transience in AGN, a new type of AGN phenomenon that has been 
 established by the ROSAT 
All-Sky Survey (RASS, Voges et al. 1999). 
X-ray transient AGN are typically identified as bright sources in 
 their 'high' state phase  during the
RASS, and are then found to be dramatically fainter or even to have
 disappeared by the time of follow-up pointed observations  years later.
Some prominent examples of ROSAT discovered transients 
are IC 3599 (Brandt et al. 1995, Grupe
et al. 1995a), WPVS007 (Grupe et al. 1995b), or  RX J0134.2--4258
(Grupe et al. 2000). 
X-ray transience has even been discovered in non-active galaxies such as 
RX J1624.9+7554 (Grupe et al. 1999b), RX J1242.6--1119 (Komossa \& Greiner
1999), or RX J1420.4+5334 (Greiner et al. 2000).

X-ray variability in AGN can be explained by either intrinsic
variability plausibly caused by such factors as changes in the
accretion rate or relativistic boosting effects (e.g. Boller et
al. 1997a) or by changes of the cold and warm absorption column
densities towards the source (e.g. Abrassart \& Czerny 2000; Schartel
et al. 1997, Komossa \& Meerschweinchen 2000).

We performed a variability study of a complete sample of bright soft X-ray
selected AGN (Grupe et al. 2001) and found RX J2217.9--5142 to be one of the
most variable sources on short as well as long timescales. Optically, we
identified RX J2217.9--5142 as a NLS1
with FWHM(\hb)=1850 \kms~ and very strong FeII emission
at a redshift of z=0.160 (Grupe et al. 1999a).

The paper is organized as follows: Sect. \ref{observe} describes the X-ray and
optical observations, Sect. \ref{results} presents the results which are discussed
in Sect. \ref{discussion}.
Throughout the paper, luminosities are calculated assuming a Hubble
constant of \mbox{$H_0 = 75$\,km\,s$^{-1}$Mpc$^{-1}$} and a deceleration
parameter of $q_0 = 0$, except if noted otherwise.
Spectral slopes, $\alpha$, 
are defined by $F\propto \nu^{-\alpha}$.

\section{\label{observe} Observations and data reduction}
RX J2217.9--5941 was observed during the RASS 
between 22 October 1990, 12:32 and 24 October 1990, 19:00
for a total of 587 s.
We used the photon event file of the RASS II processing.
Photons were extracted in a circle of $250^{''}$  radius for the source and
two circles of $400^{''}$ in scan direction for the background.  
RX J2217.9--5941 was observed twice by the 
ROSAT HRI detector: 1) ROR 702894,
06 May 1997 13:09:44 - 10 May 1997 02:12:42 for a total exosure
of 4476 s, and 
2) ROR 601124,
15 April 1998 16:32:36 - 18 April 1998 00:42:42 for a total of 9760 s. Source
photons were collected in a circle of 40$^{''}$ radius and background in a ring
with radii of 50$^{''}$-150$^{''}$ around the source.  

ASCA observed RX J2217.9--5941 on 5 May 1998.  Standard processing yielded a
total exposure in the SIS detectors of approximately 40ks. RX
J2217.9--5941 was in a low state during the observation, and analysis was
complicated by the fact that two sources, both approximately 4
arcminutes away and both detected in the HRI image 
 (Fig. \ref{hri_image}), 
were rather bright
during the observation.  These objects are identified as RX
J2217.9-5937 and RX J2218.0-5939 respectively  (sources B and C in Fig.
\ref{hri_image}).  
Thus, care was required
for imaging and spectral analysis. Therefore we discarded a period of
about 10.5 ks at the beginning of the observation that was
characterized by somewhat unstable pointing (although within the
standard limits) and a very bright background flare of unknown origin
that was inexplicably not detected by the usual background monitors.
The final usable exposure was 29.5ks.

V and B images with exposure times of 5 and 10 minutes
were taken during an identification run for ROSAT discovered X-ray sources
(Thomas et al. 1998)
on 18th August 1992 with
the ESO/MPI 2.2m telescope at La Silla.
Optical spectra were taken in 1992 and 1993  with the ESO/MPG 2.2m  
telescope, in 1995 with the ESO 1.52m,  
and with the 1.5m Ritchey-Chretien telescope at CTIO in 1998.  We note
that the 1998 observation was made with accurate spectrophotometry in
mind; therefore a 4.5$^{''}$ slit was used under photometric
conditions and all of the light from the AGN was extracted during the
reductions.  

Data reduction was performed with EXSAS (Zimmermann et al. 1998)
for the ROSAT data, MIDAS for the optical data (except for the 1998
spectrum) and XSELECT for the
ASCA data.
ROSAT HRI and ASCA count rates are converted into ROSAT PSPC count
rates using the W3PIMMS 
program of NASA's Goddard Space Flight Center
(version 2.7, 1999, http://heasarc.gsfc.nasa.gov/Tools/w3pimms.html)
assuming no spectral changes between the RASS and the pointed observations,
 and based on the power-law fit to the RASS data with the column density $N_H$
fixed to the Galactic value of Dickey \& Lockman (1990; 
$N_H$=2.58 $10^{20}$ cm$^{-2}$).

Infrared data where taken from the Infrared Processing and Analysis Center (IPAC)
using the interactive xscampi program.

\section{\label{results} Results}

\subsection{\label{ident} Identification security}
The X-ray position from the RASS is $\alpha_{2000}$=22h17m57.1s, 
$\delta_{2000}=-59^{\circ}41^{'}34.0^{''}$ with a 
2$\sigma$ error radius of 11.5$^{''}$.
Fig. \ref{v_image} displays the V image of RX J2217.9--5941 and the error radius
around the RASS position of the source. Another source is present
inside the 90\% confidence circle (\#2 in Fig. \ref{v_image}). There
are a couple of pieces of  evidence 
that \#1 is RX J2217.9--5941: 1) The position of the bright source in the center
of the HRI observation is $\alpha_{2000}$=22h17m56.7,
$\delta_{2000}=-59^{\circ}41^{'}31.2^{''}$, only 1.8$^{''}$ away from 
source \#1 in Fig. \ref{v_image} (see Table \ref{sources});
2) the identification spectrum shows a typical NLS1, which often have very steep
X-ray spectra (see Boller et al. 1996, Grupe et al. 1998a, 1999a);
3) It is the bluest object among all four bright
sources in the field (see Table \ref{sources}), and  4)
 the optical spectrum of source \#2 shows that of a K-star. 
 Using the statistic for the ratio of the X-ray to optical luminosity
of K-stars (Beuermann et al., 1999) we expect source \#2 to have a count
rate of 0.002 \cts. A $1\sigma$ deviation from the average would
result in a count rate of 0.007 \cts. So its contribution to the
X-ray luminosity of RX J2217.9--5941 is expected to be negligible.

\begin{figure}[h]
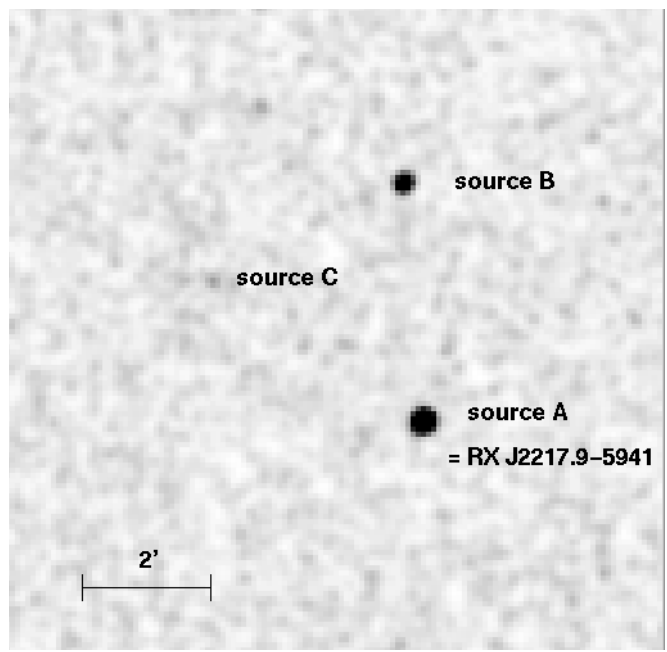

\clipfig{H2517_F1}{87}{16}{64}{196}{240}
\caption[ ]{\label{hri_image}
HRI image of RX J2217.9--5941. The sources A,B,C refere to the definitions
given in Table \ref{asca_res}. See also Sect. \ref{ident} for the exact HRI
positions of those sources.
}
\end{figure}

\begin{figure}[h]
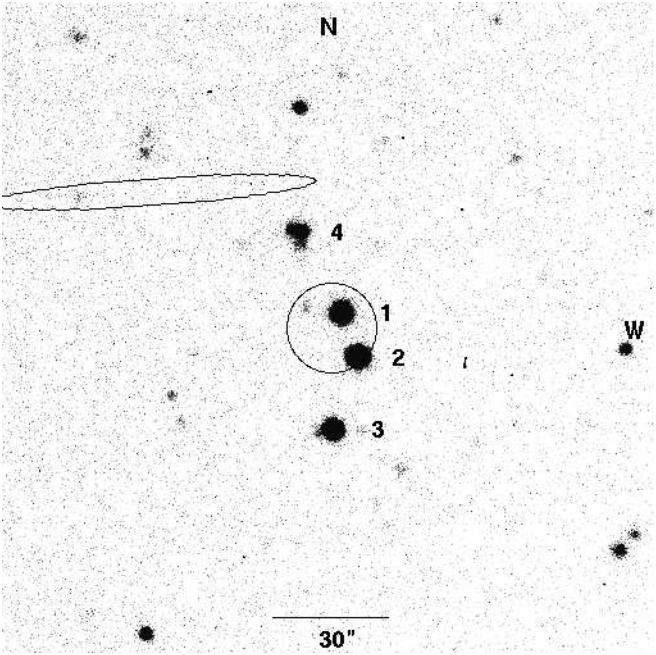

\clipfig{H2517_F2}{87}{24}{108}{183}{268}
\caption[ ]{\label{v_image}
V image of surroundings of
RX J2217.9--5941, source \#1. The center of the circle marks 
the RASS position and the radius is at the 90\% confidence level. The positions
and colours of the four sources marked here are given in Table \ref{sources}.
The IRAS position is marked by the 90\% error ellipse in the upper left part of
the Figure.
}
\end{figure}

Fig. \ref{hri_image} displays the HRI image of the surroundings of RX
J2217.9--5941. The image was created from the merged event file of both HRI
observations in order to improve the signal-to-noise ratio in that image.
The sources A, B, and C refer to the definition given in Table
\ref{asca_res}. Source A is identical with our AGN RX J2217.9--5941 and the
HRI position is given in the paragraph above. Source B has the HRI position 
$\alpha_{2000}$=22h17m55.3, $\delta_{2000}=-59^{\circ}37^{'}39.2^{''}$ and
source C $\alpha_{2000}$=22h18m03.2, $\delta_{2000}=-59^{\circ}39^{'}15.2^{''}$.
Both sources are close to the edge of the extraction radius of 250$^{''}$ of the
RASS data of RX J2217.9--5941 (source A), 
but any contribution of these sources can be neglegted.

\begin{table*}
\caption{\label{sources} Summary of the sources \#1-\#4 in Fig. \ref{v_image}
Coordinates are from the US Naval Observatory A2 scans, b and r 
are the magnitudes measured
from the USNO scans, $\rm B_j$ and OR from the APM scans,
and B, V, and B-V from the measurements of our B and V
images
}
\begin{flushleft}
\begin{tabular}{cccccccccc}
\hline\noalign{\smallskip}
& & & \multicolumn{2}{c}{USNO A2} & \multicolumn{2}{c}{APM mag} & \multicolumn{3}{c}{ESO
images} \\
\rb{\#} & \rb{$\alpha_{2000}$} & \rb{$\delta_{2000}$} & b & r & $\rm B_j$ & OR & B & V & B-V \\  
\noalign{\smallskip}\hline\noalign{\smallskip}
1 & 22 17 56.61 & --59 41 30.2 & 16.4 & 15.9 & 15.73 & 15.36 & 16.88 & 16.54 & 0.34 \\
2 & 22 17 56.06 & --59 41 41.4 & 17.3 & 15.4 & 16.17 & 14.90 & 16.98 & 16.13 & 0.85 \\
3 & 22 17 56.90 & --59 42 00.5 & 17.2 & 15.6 & 15.95 & 15.18 & 17.24 & 16.52 & 0.72 \\
4 & 22 17 58.18 & --59 41 10.0 & 17.0 & 15.4 & 16.89 & 15.69 & 18.64 & 17.65 & 0.99 \\
\noalign{\smallskip}\hline\noalign{\smallskip}
\end{tabular}
\end{flushleft}
\end{table*}

\subsection{X-rays}

\subsubsection{\label{rosat_obs} ROSAT Observations}
We found that RX J2217.9--5941 suffered a decrease in its count rate by a factor of about 30
between its RASS coverage and the HRI observation in 1998. 
During the RASS observation RX J2217.9--5941 had a count rate of 0.83$\pm$0.06
$\rm cts~s^{-1}$ (Thomas et al. 1998).
The HRI pointings result in 0.0519$\pm$0.0045 HRI $\rm cts~s^{-1}$
for the 1997 observation and 0.0054$\pm$0.0018 HRI $\rm cts~s^{-1}$ for the
1998 observation, which correspond to 0.256 and 0.026 PSPC cts$\rm~s^{-1}$ using
W3PIMMS, respectively (see Sect. \ref{observe}).

We performed standard power-law spectral fits to the RASS spectrum, with all
parameters free or with neutral absorption fixed to the Galactic value.
The results of the single power-law fits to the RASS spectra
are listed in Grupe et al. (2001). 
The best fit was obtained when the column density was free to vary;
however, this   results in less absorption than what is predicted from the Dickey
\& Lockman maps (1990). We infer this to be evidence for a soft excess
in the spectrum.  When the column density is fixed to the Galactic
value, we derived a spectral slope of  \ax=2.7, one of the steepest
among the AGN in 
our soft X-ray selected AGN sample (Grupe et al. 2001). The rest-frame
luminosity in the ROSAT band is log $L_{0.2-2.0}$=37.5 [W].
Due to the small number of collected photons during the
RASS coverage complicated spectral fits were not warranted.  

We tried to use the poor energy resolution of the HRI and derived a hardness 
ratio as
defined by Hu\'elamo et al. 2000. However, the gain of the HRI was adjusted in
December 1997, between our two observations. Therefore, the systematic errors caused by
this gain shift and gain adjustment are too large to retrieve meaningful 
results (e.g.\ Leighly et al.\ 1997).

\begin{figure}[h]
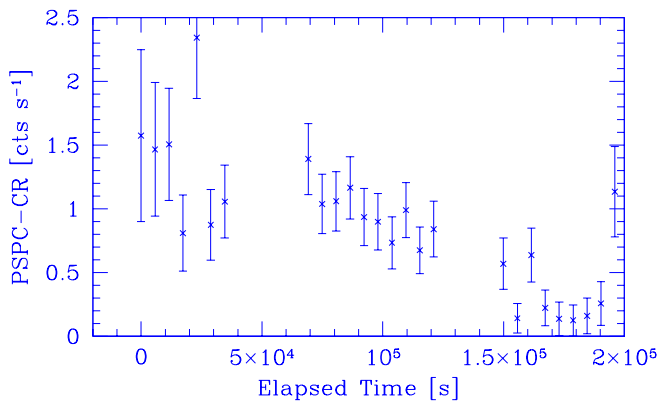

\clipfig{H2517_F3}{87}{17}{145}{169 }{240}
\caption[ ]{\label{lightcurve}
Light curve of RX J2217.9--5941 during the RASS coverage. 
}
\end{figure}

\begin{figure}[h]
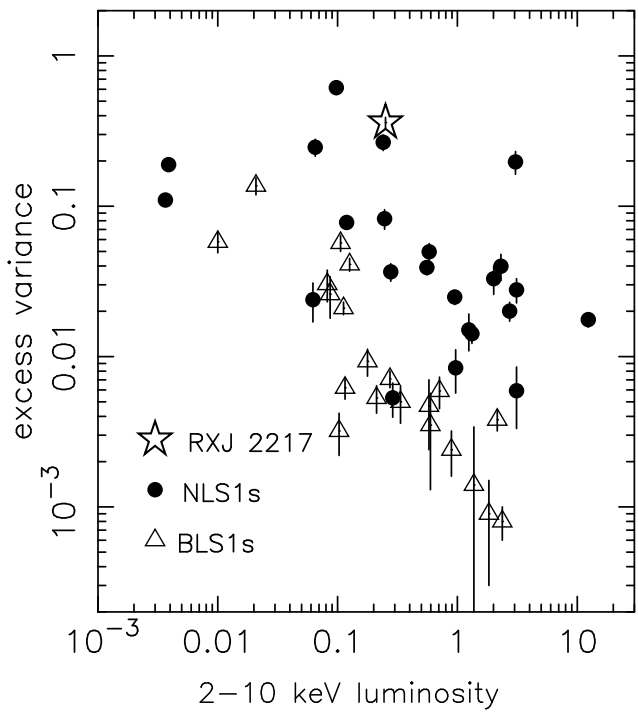

\clipfig{H2517_F4}{87}{25}{45}{170}{195}
\hspace*{2.5cm}
\clipfig{H2517_F4}{43}{40}{30}{120}{40}
\caption[ ]{\label{variance}
Excess variance of RX J2217.9--5941 vs. 2-10 KeV 
rest-frame luminosity in comparison with the
sample of Leighly (1999a,b). The 2-10 keV rest-frame luminosity is given in units
of $10^{37}$ Watts. 
}
\end{figure}

Fig. \ref{lightcurve} displays the RASS lightcurve. Remarkably, the
count rate decreased by a factor of $\approx$12 in about two days.  We
could not detect any significant spectral change between the high and
the low states due to the low photon statistics in the RASS data.  A
variability test of the RASS light curve gave a $\chi^2/\nu$=4.2 which
makes RX J2217.9--5941 one of the most variable sources on short
timescales throughout our complete sample of soft X-ray selected AGN
(Grupe et al. 2001).  A more common parameter often used in
variability studies is the excess variance (e.g.  Nandra et al. 1997,
Leighly 1999a). The excess variance of RX J2217.9--5941 is
$\sigma_{\rm rms}^2$=0.36\pl0.03.  In order to compare with the {\it
ASCA} results on NLS1s, we estimated the 2--10 keV rest-frame
luminosity using a two component spectral fit to the RASS data
assuming the ASCA spectral slope of \ax=1.6 (Section 3.2.2).  We find
that with the 2--10 keV rest-frame luminosity log $L_{2-10}$=36.4, RX
J2217.9--5941 is one of the more variable NLS1 (Fig.  \ref{variance},
see also Fig. 3 in Leighly 1999a).  Note that the length of the RASS
observation is similar to that of the ASCA observations reported in
Leighly 1999a, so that the comparison with the results reported in
that paper is valid, but also note that in general the excess variance
observed in hard X-rays is lower than observed in soft X-rays (e.g.\
Leighly 1999a).  The HRI light curves were checked for
variability as well. In both observations the result from the $\chi^2$ test
is consistent with no variability.

\subsubsection{ASCA Observations}

Fig. \ref{asca-ima}
shows the ASCA soft and hard band images generated 
from the
SIS0 and SIS1 detectors.  These have been adaptively smoothed, divided
by an exposure map generated by the {\it ascaexpo} tool, and further
smoothed with a boxcar average function 5 pixels wide. The 'x' and '+'
mark the HRI positions of RX J2217.9--5941 
 (source A in Table \ref{asca_res} and Fig. \ref{hri_image})
and the bright unknown source RX
J2217.9--5937  (source B), respectively.  This demonstrates that the photometry in
the ASCA image is adequate to identify the rather faint source as RX
J2217.9--5937.  RX J2217.9--5941 is clearly detected in the soft-band image but
cannot be clearly identified in the hard band image. 

Analysis of the ASCA data is complicated by the two sources only 4$^{'}$
away.  While the point-spread function of ASCA near the optical axis
of the SIS has a relatively sharp core about 1 arcminute FWHM, it has
broad wings (half-power diameter of 3 arcminutes), and it is a
function of energy.  Thus the sources near RX J2217.9--5941 contaminate the
spectrum drawn from a region including that source in a way that is
non-negligible and difficult to estimate. It is possible to extract useful and
unambiguous spectral information from faint sources located near
bright sources by using the TRACE\_ASCA software developed by A. Ptak
to simulate the contribution of the bright source in the extraction
region of the fainter source.  However, we do not employ this rather
complex analysis technique here simply because the statistics are
rather poor and are likely to dominate the spectral analysis.  The GIS
detector smears the X-ray image more and therefore we do not consider
the GIS data at all here.

We finally decided to extract and analyze spectra from a region
obtained within a 1.5 arcminute radius region.  This region size is
much smaller than the nominal SIS extraction size of 4 arcminutes
radius; however, the proximity of nearby sources required small region
sizes so that the regions would not overlap and so that the
contamination of the flux from the neighboring sources would be
minimized.  Significant source photons will be outside of these small
region sizes; however, the ancillary response files account for the
reduction in effective area and thus the measured flux would not be
affected.  In this case, however, the measured flux will be an upper
estimate because of the contamination by neighboring sources.  The
ASCA PSF is energy dependent and hard photons are scattered more than
soft photons. This effect is also accounted for by the ancillary
response file, except for the fact that the high energy spectrum will
be less well constrained than it would be if the nominal region size
could be used.

\begin{figure*}[h]
\clipfig{H2517_F5}{170}{10}{97}{205}{180}
\caption[ ]{\label{asca-ima}
ASCA image of RX J2217.9--5941. The HRI position of RX J2217.9--5941 
 (source A in Table \ref{asca_res} and Fig. \ref{hri_image})
is marked
by an `X' and the HRI position of the bright source RX J2217.9-5137 (source B)
by
a `+'. Note that RX J2217.9--5941 is only clearly detected in the soft
band image, a fact that supports the steep spectrum derived from
spectral analysis. HRI positions for both sources are given in Sect.
\ref{ident}.
}
\end{figure*}

We first attempted to determine whether the sources were variable.
Light curves binned by orbit were extracted and examined, and no
clearly convincing evidence of variability was found during the
observation in any of the three sources.  Statistical evaluation of
the variability was hampered by the low count rates: the number of
photons per orbit was so small that the uncertainties should be
governed by Poisson rather than Gaussian statistics and thus the
errors were not trivially evaluated.  Maximum likelihood techniques
can be applied when the count rates are low (e.g. Leighly 1999a);
however, these techniques must be used on counts light curves rather
than rate light curves, and counts light curves could not be
constructed because there was a wide distribution of orbital net
exposures.

Table \ref{asca_res} gives the results of spectral fitting.
Statistics were poor enough that the only model used to fit the
spectra was a power law with the column density fixed at the Galactic
value.  It is noteworthy that the best fitting spectral index of RX
J2217.9--5941 is \ax=1.6; this is the typical maximum value obtained
from other NLS1s (e.g.\ Leighly 1999b).  We note that because of the
source contamination and because of the small number of photons, our
derived ASCA spectral index must be considered extremely tenative.
However, it seems fairly clear that the spectrum of the object is not
flat as it would be if the spectrum were either dominated by
reflection ($\alpha_{\rm X} \le 0$; e.g.\ Matt et al.\ 1996) or
absorbed.  The fact that the source is detected clearly only in the
soft band image also supports a soft spectrum.

\begin{table*}
\caption{\label{asca_res} Power-law fits to the ASCA spectra of the objects
listed below. The absorption parameter $N_H$ was fixed to the galactic value
(see text). The 2-10 keV model flux is given in units 
of $\rm 10^{-17} W~m^{-2}$, the predicted and measured 
HRI count rates are in units of $\rm 10^{-3}~cts~s^{-1}$. 
}
\begin{flushleft}
\begin{tabular}{clccrcccc}
\hline\noalign{\smallskip}
& & total \# & & 2-10 keV & & HRI predicted & \multicolumn{2}{c}{HRI measured CR} 
\\
\rb{Source} & \rb{Object$^{a}$}  & of photons$^{b}$ &  \rb{$\alpha_{\rm X}$} 
& Flux$^{c}$ &
 \rb{$\chi^{2}$/dof} & count rate$^d$ & 1997 & 1998 \\
\noalign{\smallskip}\hline\noalign{\smallskip}
A & RX J2217.9--5941  & 187 &    $1.60^{+0.66}_{-0.58}$ & $2.7^{+2.5}_{-1.6}$ &
6.2/11 & $2.88_{-1.10}^{+1.54}$ & 51.9\pl4.50 & 5.40\pl1.80 \\ 
\noalign{\smallskip}
B & RX J2217.9--5937  & 284 &    $1.09^{+0.32}_{-0.29}$ & $9.7^{+3.5}_{-3.2}$ & 
21.8/20 & $3.92_{-0.88}^{+0.96}$ & 8.37\pl2.36 & 4.36\pl1.24 \\ 
\noalign{\smallskip}
C & RX J2218.0--5939$^e$ & 126 & $0.69^{+0.52}_{-0.52}$ & $13.2^{+9.0}_{-6.3}$ & 
5.5/6 & $2.65_{-0.86}^{+1.10}$ & --- & ---  \\
\noalign{\smallskip}\hline\noalign{\smallskip}\\
\end{tabular}

$^a$ The exact HRI positions of these sources are given in Sect. \ref{ident}. \\

$^b$ Total number of photons in the extraction regions for SIS0+SIS1
between 0.6 and 5 keV.  This includes source photons, background
photons, and photons from the PSF wings of the other sources.
The number of contributing background photons is estimated to
be between 80 and 100 counts in the same energy band for two
detectors.  \\

$^c$ Flux uncertainties were generated by determining fluxes for values of
photon index and normalization separately that produced worse fits by
$\Delta\chi^2$ of 2.71.

$^d$ Based on spectral fits to the ASCA data and usage of the response matrix.

$^e$ Note that RX J2218.0-5939 was observed only in the SIS0 detector.

\end{flushleft}
\end{table*}

The identity of the two other objects in the ASCA field of view is
unknown.  The HRI astrometry is good enough that RX J2217.9$-$5937 (source B)
can
be fairly confidently identified with an object appearing in the
Digitized Sky Survey at 22 17 59.2, $-$59 37 41 (J2000).  Although the
USNO catalog magnitudes are only reliable within about 0.5 magnitude,
we can use them to get an idea of the identity of the object.  The
USNO catalog lists the R and B band magnitudes of this source as 18.0
and 18.7 respectively.  Assuming that the spectrum is a power law
gives an estimate of the V band magnitude of 18.4.  The best fitting
model of the ASCA spectra gives an estimated flux in the 0.3--3.5 keV
band of $1.5 \times 10^{-17}\rm \, W\,m^{-2}$.  Comparison
of these values with the X-ray source nomogram (Maccacaro et
al. 1988), and assuming no intrinsic absorption or reddening, shows
that the optical and X-ray properties of the object are consistent
with it being an AGN.

Identification of the optical object associated with 
RX J2218.0-5939  (source C)
is more difficult because it is a weaker object.  There is an
optical source within 7 arcseconds identified in the USNO catalog as
having R and B magnitudes of 18.3 and 20.9 respectively.

\subsection{Optical}

\begin{figure}[t]
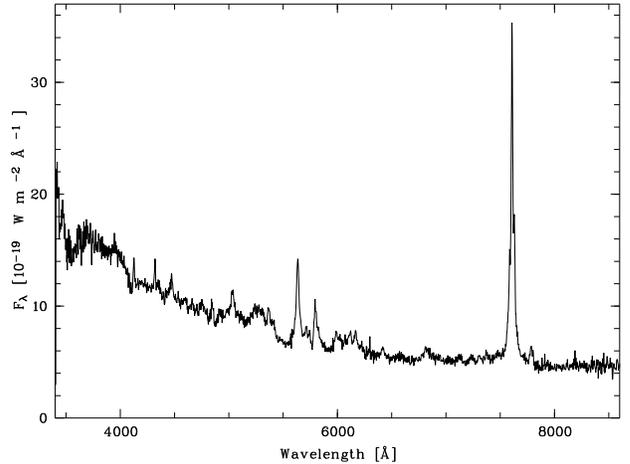

\clipfig{H2517_F6}{87}{18}{10}{273}{195}
\caption[ ]{\label{opt_spec}
Optical spectrum of RX J2217.9-5941 obtained with the 1.5m telescope at CTIO
}
\end{figure}

The optical spectrum of RX J2217.9--5142 (Fig. \ref{opt_spec})
shows the typical spectrum of a
Narrow-Line Seyfert 1 galaxy. The
full width of the H$\beta$ line was measured to 1850\pl100 $\rm km~s^{-1}$. 
The spectrum appears to be unreddened (H$\alpha$/H$\beta$=3.4,
\aopt=0.3; Grupe et al. 1999a).
We subtracted an FeII template from all optical spectra of RX J2217.9--5941 
(see e.g. Boroson \& Green 1992, Grupe et al. 1999a). 

The redshift of RX J2217.9--5941 based on the H$\alpha$ and \hb~ lines
was z=0.160 (Grupe et al. 1998a, 1999a). The [OIII] lines show a
blueshift of about 570\pl20 \kms with respect to the Balmer lines;
this will be discussed further in Grupe \& Leighly (in prep.).
The line
width of the [OIII] lines is extremely broad compared to the rest of
the sample. We measured a FWHM([OIII]$\lambda$5007)=1075\pl150 \kms
(Grupe et al. 1999a).

Erkens et al. (1997) reported a correlation between the strength of
coronal iron lines and the steepness of the ROSAT spectra in
AGN. Because RX J2217.9--5941 has a very steep X-ray spectrum, we
looked for coronal iron lines but did not find any in any of the
spectra.  A comparison of the spectra taken in 1992, 1993, 1995, and
1998 does not show detectable changes in line and continuum fluxes.

\subsection{Other wavelength ranges}
{\em Infrared:}
RX J2217.9--5841 is listed as a bright
infrared IRAS source in our soft X-ray selected AGN sample, and it is
also listed as in infrared source in the  samples of Boller et al. (1992, 1997b)
and Moran et al. (1996).  However, a comparison of the IRAS and ROSAT
positions reveals that  RX
J2217.9--5941 is probably not the infrared source. The IRAS position marked in Fig.
\ref{v_image} is
$\alpha_{2000}$=22h18m03.3 and $\delta_{2000}=-59^{\circ}40^{'}58^{''}$ (Boller
et al. 1997b). So either the faint object 10$^{''}$ 
north of the IRAS position or
object  \#4 in Fig. \ref{v_image} is the optical counterpart of the IRAS source. 

{\em Radio:} We have searched for radio data for RX J2217.9--5941. The
nearest entry in the Parkes-MIT-NRAO survey catalog (PMN, Wright et
al. 1994, Condon et al. 1998) is at $\alpha_{2000}$=22h17m28.1s,
$\delta_{2000}=-59^{\circ}31^{'}10.0^{''}$, about 11$^{'}$ away from
the position of our source, and too far away to be associated with RX
J2217.9--5941.

\subsection{\label{sed} Spectral Energy Distribution}
\begin{figure}[h]
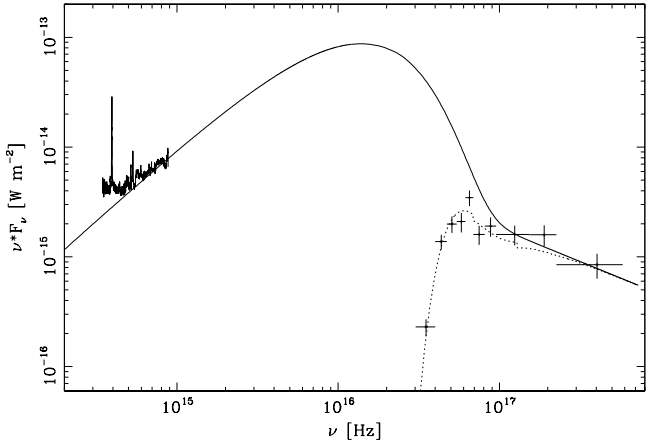

\clipfig{H2517_F7}{87}{37}{31}{255}{180}
\caption[ ]{\label{rxj2217_sed}
Optical to X-ray spectrum of RX J2217.9--5941. The solid line shows the model as
described in Sect. \ref{sed}, the dashed line the absorbed spectrum fitted to
the RASS spectrum (marked by crosses).   
}
\end{figure}

In order to obtain an estimate of the bolometric luminosity, which is
used 
later to obtain a limit on the black hole mass,
we performed a two-component
spectral fit to the optical and RASS X-ray
 data consisting of a low-energy power law with exponential 
cut-off  and a high-energy power law. The low-energy part
represents a simple geometrically thin accretion disk model
in LTE (e.g. Krolik 1999),
with $\alpha$=--1/3. 
 The optical spectrum 
(see Fig. \ref{opt_spec}) was corrected for
interstellar extinction using the relation between $\rm E_{B-V}$
 and Galactic column  density
given by Diplas \& Savage (1994), a value of $\rm A_V/E_{B-V}$ = 3.2 
(Seaton, 1979), and the wavelength dependence of $\rm A_{\lambda}$
 derived by Cardelli et al. (1989). At the blue  end of
the optical spectrum ($\nu~ =~8.4~ 10^{14}$ Hz), where the contribution of the 
underlying galaxy
is small, we obtained a flux of $\nu F_{\nu}~ =~ 7.40~ 10^{-15}$ W m$^{-2}$. 
We fixed the column density
 $N_H$ to the Galactic value and the hard X-ray spectral slope to
\ax=1.6, the value derived from the ASCA data (see Section 3.2.2 and Table \ref{asca_res}).
The fit to the X-ray data then gave a cut-off energy of 41.9\pl1.2 eV, and a 
normalisation for
the high-energy power law of $\nu  F_{\nu}~ =~ (1.06$\pl$0.26)~ 10^{-15}$ 
W m$^{-2}$
 at 1 keV, with $\chi^2/\nu$ = 10.3/9. The Wien tail of the low-energy 
 component (in the X-ray regime) corresponds
to a power law with \ax = 4.4. 
Fig. \ref{rxj2217_sed} displays the result of this fit. Although the optical
spectrum obtained in 1998 was not simultaneous with the RASS data, V magnitudes 
obtained in Aug. 92
and Sept. 93 differ by 0.08 mags only.  
 The total luminosity over the optical to X-ray range, where most of the
energy is radiated, is log L = 39.1 [W].

Based on the model described above we estimated the rest-frame
fluxes at 2500\AA~ and 2.0
keV in order to derive the optical/UV to X-ray slope \aox. For the RASS
observation we found \aox=1.52, and for the two pointed HRI observations in 1997
and 1998 we derived \aox=1.62 and \aox=2.00, respectively, assuming that the
spectral shape had not changed and that the optical flux at 2500 \AA~
was constant. This is in the range reported for the PG quasar 
sample of Tananbaum et al. (1986) by Wilkes et al. (1994; see their Fig. 6b).
Following Eq. 12 in Wilkes et al. (1994), from the luminosity at 2500\AA~ an
\aox=1.47 would be expected. Here a H$_0$=50 $\rm km~s^{-1}~Mpc^{-1}$ was used
to be consistent with Wilkes et al. 1994.
This is in agreement with the RASS value. 
However, the second HRI and the ASCA observations lay at the very steep end
compared to the PG quasar sample.   We also compared our values with
the larger data base for radio-quiet quasars compiled by  Yuan et
al. (1998).  They find a mean \aox~ of 1.51\pl0.03 for the 
redshift and luminosity range of RX J2217.9--5941. This is precisely
the value we
found for the RASS observation, but again much smaller than the values we
obtained from the HRI and ASCA observations.

We also checked log($f_x/f_o$) as given in Eq. 2 in Beuermann et al.\
(1999) and compared it with their Fig.\ 2. We found that the RASS as
well as the first HRI pointing still fall into the region where AGN
are expected. The second HRI pointing is, considering the spectrum is
still steep, way off the AGN region (log($f_x/f_o$)= +0.54, +0.03, and
--0.97 for the RASS, the 1997 HRI and the 1998 HRI observations,
respectively). 

\section{\label{discussion} Discussion}

RX J2217.9--5941 is one of the most variable AGN on short timescales of
days to long timescales of years in our sample of soft X-ray selected AGN
(Grupe et al. 2001). 
The variability of RX J2217.9--5941 shows:
\begin{enumerate}
\item a strong decrease in PSPC count rate in its `high' state during the RASS,
but no significant variability during the three observed `low' states during the
pointed ROSAT and ASCA observations.
\item an apparent slow decrease in count rate/flux on long time scales.
\end{enumerate}

In principle, variability can have two origins:
either it is intrinsic or
it results from changes of the cold or warm absorption column densities. 
In the first case, the variability originates in the AGN engine
itself, and it may be either persistent or a  
transient.
In the second case, absorbing clouds are in the line-of-sight.   
We now discuss the possibilities of a transient event or persistent variability
in turn.

\subsection{Transient event?}

Some results  seem to
argue for a transient nature of RX J2217.9--5941: 
\begin{enumerate}
\item During three pointings after the RASS the source was seen in `low' state.
\item From the HRI pointing in 1997 to the HRI observation in 1998 the source
decreased in count rate by a factor of about 10 and even the ASCA observation
three weeks after the 1998 HRI observation is fainter by about a factor of 2
(see Table \ref{asca_res}).
\item The X-ray spectrum is steep. All transients discovered so far showed very
steep X-ray spectra.   
\end{enumerate} 
We can imagine two forms of X-ray outburst events, either an
 instability in the accretion disk that causes an dramatic increase in
 the accretion rate, or a tidal disruption of a star by the central
 black hole.  An instability in the accretion disk can cause a higher
 accretion rate over a short time or a decrease in the accretion
 rate. In both cases we would expect to see a decrease in the X-ray
 flux over long time-scales. This can be a one-time event, but it can
 be also persistent variability.  Tidal disruption of a star by a
 central black hole appears when the path of a star comes too close to
 the central black hole and gets disrupted by the gravitational force
 (Rees 1990, Ayal et al.\ 2000).  This model can explain X-ray
 outbursts in non-active galaxies such as RX J1624.9+7554 (Grupe et
 al. 1999b) or the Seyfert 2 galaxy IC 3599 (Brandt et al.\ 1995,
 Grupe et al.\ 1995). However, all those objects are inferred to have
 black hole masses in the order of $10^6$ \msun. The black hole in RX
 J2217.9--5941 is about two orders of magnitude larger.  
  We can estimate the lower limit of the black hole mass by
assuming that the highest luminosity seen in this source is the
Eddington luminosity. With a mean bolometric luminosity of log $L_{\rm
bol}$ = 39.1 [W] derived from the RASS and the optical observations
(see Sect. \ref{rosat_obs}) we get a minimum black hole mass of RX
J2217.9--5941 of $\approx 10^{8}$\msun. 
Even if it is
 not impossible for a 10$^8$\msun~ black hole to disrupt a star, it is
 less likely simply due to the size of the black hole.  The star must
 cross the region between the event horizon and the tidal disruption
 radius in order for us to see a transient event.  This region becomes
 smaller as the BH mass increases because the radius of the event
 horizon scales with the mass of the black hole while the tidal
 disruption radius scales with $M^{1/3}$ (Rees 1990).

\subsection{Persistent variability?}

The variable X-ray source could be the accretion disk itself. A
cut-off powerlaw, which has a shape of a simple accretion disk model,
fits the observed data in the optical and X-ray part of the spectrum
quite well.  Thus the X-rays are emitted from the inner part of the
disk (3-10 $\rm R_S$), while the optical emission is produced further
out.   The bolometric luminosity of log $L_{bol}$ = 39.1
observed in RX J2217.9--5941 requires an accretion rate of 1.4 \msun
$\rm y^{-1}$ assuming an efficiency of $\eta$=0.1 (L=$\eta \dot
M$c$^2$).  This high accretion rate might explain the strong decrease
in count rate during the `high' state RASS observation while during
the `normal' low states in later pointings the accretion rate is more
stable.  That is, during the RASS, the source may have been accreting
close to the Eddington limit, giving a very soft X-ray spectrum
(e.g. Pounds et al. 1995, Ross et al. 1992).  In later years, the
accretion rate decreased, resulting in a lower temperature in the
inner part of the disk and shifting the soft X-ray component out of
the ROSAT window; this explanation was also proposed for the X-ray
transient WPVS007 (Grupe et al. 1995b).  

However, it can not be the soft X-ray part alone that must have
changed.  During the RASS, the contribution to the total count rate
from the hard x-ray spectrum alone was 0.28 \cts, assuming \ax=1.6.
Assuming that during the pointed observations all of the soft
component flux was gone, the HRI count rate predicts 0.155 PSPC \cts~
for the 1997 HRI observation and 0.016 PSPC \cts~ (with \ax=1.6 and
$N_H$= galactic).

An alternative explanation for persistent variability is relativistic
Doppler boosting. We cannot completely exclude this explanation,
however, the boosting factor is very sensitive to the inclination
angle and strongest when the disk is viewed edge-on. RX J2217.9--5941
is a NLS1 and these sources are usually thought to be viewed at low
inclination angles (e.g.\ Grupe et al.\ 1998b)
and therefore the boosting factor should be low.  Nevertheless, Boller
et al. (1997) suggested relativistic Doppler boosting to explain the
rapid and persistent variability in the NLS1 IRAS13224--3809. 
However, we observe strongly blueshifted
broad high-ionization lines with nearly no emission redward of the rest
wavelength, while the low ionization lines are narrow and symmetric.
A disk-wind model most naturally explains these results, and therefore
an edge-on orientation is very unlikely (Leighly \& Halpern 2001;
Leighly 2000).

Also arguing for a non-transient nature of RX J2217.9--5941 is the
optical-to-X-ray slope \aox. During the RASS it was in the range
found for quasars by Yuan et al.\ 1998, Wilkes et al.\ 1994, and
Brandt et al.\ 2000, and agrees with the expected value of \aox~
derived from the optical/UV luminosity as defined by Wilkes et al.\
1994.  The slope \aox~ derived from the HRI and ASCA observations are
much steeper compared to those samples, indicating that RX
J2217.9$-$5941 is deficient in X-rays.  Brandt et al.\ (2000) have
investigated ``soft X-ray weak'' AGN and find a clear correlation
between the CIV absorption line equivalent width and \aox, suggesting
that soft X-ray weak objects are absorbed (also Schartel et al.\
1997). There was certainly no X-ray absorption present during the RASS
observation (when the \aox was in the normal range).  Although the
ASCA results are extremely tentative because of poor statistics and
contamination, we find no evidence to suggest that the X-ray spectrum
is absorbed when RX J2217.9$-$5941 was in the faint state.  However, we
will have to wait for follow-up Chandra or XMM observations to make
any definitive statements about the shape of the X-ray spectrum and
whether or not there is any X-ray absorption present.  HST
observations would reveal whether there are deep UV absorption lines
present. 

 Another argument for persistent variability in RX
J2217.9--5941 is that strong variability is common among NLS1 (Leighly
1999a,b) and RX J2217.9--5941 is a typical NLS1.

Another possibility is that we are looking at two kinds of
variability.  Absorption could have been responsible for the
short-time variability during the RASS when the source was in its
`high' state, when there was a strong decrease with no evidence for a
change in spectral shape (although due to the poor signal to noise, we
cannot place very strong constraints on the spectral variability).  If
there were no spectral variability, the source flux variability could
have been caused by a cold absorbing cloud moving over the a period of
two days into the line of sight between us and the X-ray source.  If
the central region is filled with patchy clouds, they can indeed cause
variability and has been proposed by several workers (e.g.  Abrassart
\& Czerny 2000; Schartel et al., 1997; Komossa \& Meerschweinchen
2000).

\section{Summary and Conclusions}
\begin{enumerate}
\item RX J2217.9--5941 has shown an remarkable decrease in count rate by a
factor of  about 40 between the RASS and pointed observations 8 years later.
\item During the 2 day RASS coverage the count rate decreased by a factor of 15.
\item The optical/UV to X-ray spectral slope \aox~ and the
flux ratio $f_X/f_o$ argue more for a persistent variability than a
one time outburst event.
\item In case of an outburst event, tidal disruption of a star by the central
black hole is most unlikely due to the mass of the black hole of at least $10^8$
\msun.
\item The [OIII] line system is blueshifted with respect to the \hb~line and
the [OIII] emission lines are broader (FWHM([OIII])=1075\kms) than the
\hb~emission line.
\item The only way to
see if RX J2217.9--5941 is a transient source with a constantly
decreasing count rate
or if it is indeed persistent variability, is to monitor this source
with new X-ray missions like XMM and Chandra. With XMM and its Optical Monitor
we would even be able to observe in the optical/UV and X-ray ranges 
simultaneously. 
\item A possible check to estimate the ``normal'' \aox~ is to observe the CIV
absorption line in the UV.
\end{enumerate}

\acknowledgements{ 
We would like 
to thank Prof.\ Jules Halpern for taking the optical spectrum of RX
J2217.9--5941 at CTIO in 1998 and Dr. Klaus Reinsch for observing source \#2
(see Table \ref{sources}) at ESO in January 2001.
We thank Drs.\ Wolfgang Brinkmann and 
 Mario Gliozzi for their comments and suggestion on the manuscript.
This research has made use of the NASA/IPAC Extragalactic
Database (NED) which is operated by the Jet Propulsion Laboratory, Caltech,
under contract with the National Aeronautics and Space Administration. 
We also
used the IRAS data request of the Infrared Processing and Analysis Center 
(IPAC), Caltech. KML gratefully acknowledges support through NAG5-7971 (NASA,
LTSA).
The ROSAT project is supported by the Bundesministerium f\"ur Bildung
und  Forschung (BMBF/DLR) and the Max-Planck-Society.

This paper can be retrieved via WWW: \\
http://www.xray.mpe.mpg.de/$\sim$dgrupe/research/refereed.html}

   \end{document}